\documentclass[12pt]{article}

\usepackage{mathrsfs,amsbsy,amssymb,latexsym,amsfonts,amsmath,cite}

\global\arraycolsep=3pt
\def\ideq{\equiv}               % triple equal sign
\def\expectation{\widehat}

\begin{document}

~
\vspace{20mm}

\begin{center}
{\Large 
An iterated search for influence from the future on the Large Hadron Collider
}

\vspace{20mm}

%%% {\large
%%% Adilah Hussein}
%%% \\
%%% {\it
%%% Frankfurt Institute for Advanced Studies,
%%% Johann Wolfgang Goethe University,
%%% \\
%%% Frankfurt, Germany}\footnote{hussein@fias.uni-frankfurt.de}
%%% \\
%%% and
%%% \\
{\large
Iain Stewart}
\\
{\it
Dept. of Computing,
Imperial College of Science, Technology and Medicine,
\\
London, U.K.}\footnote{ids@doc.ic.ac.uk}
\end{center}

\vfill

\begin{abstract}
We analyse an iterated version of Nielsen and Ninomiya (N\&N)'s
proposed card game experiment to search for a specific type of
backward causation on the running of the Large Hadron Collider (LHC) at CERN.
We distinguish ``endogenous" and ``exogenous" potential causes of
failure of LHC and we discover a curious ``cross-talk" between
their respective probabilities and occurrence timescales when
N\&N-style backward causation is in effect.
Finally, we note a kind of ``statistical cosmic censorship"
preventing the influence from the future from
showing up in a statistical analysis of the iterated runs.
\end{abstract}

\bigskip
\newpage

\section{Introduction}
Nielsen and Ninomiya \cite{hep-ph/0612032} recently proposed a
particle physics model with the property that probabilities for events in the
near future (say time $t_1$) of an ``initial" state (say time $t_0$)
depend globally on the action for complete
spacetime histories, including the parts of them further in the future
than $t_1$.
The usual simplification, where we in practice consider (and sum over)
only the parts of histories between spatial hypersurfaces $t_0$ and $t_1$,
does not apply.
This gives rise to a form of backward causation.
Things like branching ratios for events here and now can depend on
the ways the various alternatives can be continued to later times.
Events involving Higgs production,
such as the running of the Large Hadron Collider (LHC) at CERN,
are the leading candidates for such effects in their model.

In \cite{0707.1919}, Nielsen and Ninomiya (henceforth N\&N)
propose a card game experiment where influence from the future
upon the running of LHC could affect the odds of drawing
cards with various instructions, such as ``run LHC normally" or
``shut down LHC".

The card game version analysed by N\&N in quantitative detail
is a simple one-shot affair where ``shut down LHC" and
``run LHC normally" are present in mixing ratio $p$ to $1-p$.
This protocol, while providing maximal contrast between the two branches,
does require a huge level of ``community self-discipline" which
seems unlikely to be achieved in practice.
Upon the drawing of ``shut down LHC" the community would likely
take the line that ``we didn't really mean it" and proceed with
the planned runs. That very fact -- a feature of the potential time evolution
further into the future (running LHC) than a
particular event (drawing a card) -- would infect the detailed structure of
the influence from the future upon the present-day enactment of the protocol.

In this paper we analyse an iterated version of the N\&N protocol.
The same $p : 1-p$ pack as just described is used, but a new draw
is performed at the start of each agreed time period
({\it e.g.} day, week, month)
and the instructions
``run LHC normally" or ``shut down LHC" refer just to that time period.
Thus at no stage is the overwhelming demand being made that
LHC be permanently closed.
Further, the pattern of runs and idle periods thrown up by enacting
the iterated protocol
is a ``deliverable" in its own right, no less than the LHC run data,
and available for statistical scrutiny by interested parties.

\section{The relevant aspects of the N\&N model}
In \cite{hep-ph/0612032}, N\&N discuss the possible impact on the usual
classical or quantum action rules -- respectively, find a history with
extremal action, or sum over the exponentiated actions of all histories --
of allowing an {\em imaginary part} in the Lagrangian specifying a history's
action. They give reasons for the Higgs field to be the place one could
expect an imaginary part to show up.

The mathematical effects of allowing an imaginary part in the Lagrangian
are subtle and far-reaching, and according to \cite{hep-ph/0612032}
may even extend as far as determining
(with probability very close to 1) a {\em particular} solution to the equations
of motion, rather than just a recipe for computing some features of a
solution from other features.
Stepping back from these lofty longer-term goals, though,
N\&N in \cite{0707.1919} specialize the discussion
to a simplified case of their full model,
where roughly identifiable ``classical" trajectories
({\it i.e.} with probabilities rather than amplitudes attached)
have been located as solutions, and the influence of the imaginary part
of the Lagrangian is a simple multiplicative one, as follows.

Consider two solutions (``classical" histories) which,
in the {\em absence} of an imaginary part in the Lagrangian,
would have been assigned equal probabilities.
Suppose that on one history a machine such as LHC is built and
produces Higgs particles, while on the other history no such machine
is built.
Then, in the {\em presence} of the imaginary part of the Lagrangian,
the relative probabilities of the two solutions are modified
by an exponential in the number of Higgses:
\begin{eqnarray}\label{N&N_multiplicative_influence}
\frac{P(\mathrm{sol.}_\mathrm{with~machine})}
     {P(\mathrm{sol.}_\mathrm{without~machine})}
=
C^{\sharp\mathrm{Higgses}}
\end{eqnarray}

(Of course Higgs particles can be produced without deliberate intent,
for example in the hot big bang. In \cite{hep-ph/0612032}, N\&N give
reasons for believing that at least in the classical approximation
under consideration, the early universe may be taken as fixed across the
branches, and it is legitimate to focus on the differences in
{\em contemporary} Higgs production between one branch and another.
We will accept their argument for the purposes of this paper.)

Although it is not stated explicitly in \cite{0707.1919},
we take it that the formula (\ref{N&N_multiplicative_influence})
extends in the obvious way to solutions with unequal probabilities.
That is, using ``uninfluenced" and ``influenced" to refer to
probabilities in the absence and in the presence respectively
of a Lagrangian imaginary term, we have:
\begin{eqnarray}\label{extended_N&N_multiplicative_influence}
\frac{P_\mathrm{influenced}(\mathrm{sol.}_\mathrm{with~machine})}
     {P_\mathrm{influenced}(\mathrm{sol.}_\mathrm{without~machine})}
=
\frac{P_\mathrm{uninfluenced}(\mathrm{sol.}_\mathrm{with~machine})}
     {P_\mathrm{uninfluenced}(\mathrm{sol.}_\mathrm{without~machine})}
C^{\sharp\mathrm{Higgses}}
\end{eqnarray}

In (\ref{N&N_multiplicative_influence})
or (\ref{extended_N&N_multiplicative_influence}) above,
if $C < 1$ Higgs production is suppressed in processes generally,
while if $C > 1$ it is enhanced.
N\&N consider the latter case effectively ruled out by the lack of
apparent conspiracies of coincidence all around us producing
more than their normal fair share of Higgs particles.
Thus they (and we) focus on the case $C < 1$.

\subsection*{Normalization}
A collection of mutually exclusive and exhaustive classical histories
should be assigned probabilities which sum to 1.
In standard prescriptions for assigning probabilities to classical histories,
such as multiplying out the branching factors for each path on a
tree of alternatives, this happens automatically provided simple local
constraints are obeyed
({\it e.g.} that the branching factors sum to 1 at each node,
in the tree of alternatives case).
No final global normalization is required.

The simplified N\&N model with its multiplicative influence
(\ref{N&N_multiplicative_influence}) or
(\ref{extended_N&N_multiplicative_influence})
unfortunately does not assign
an individually computed probability to each history.
Only ratios of probabilities are prescribed.
To achieve these ratios
one first computes the {\em uninfluenced} probabilities by the
usual formal or informal classical reasoning ({\it e.g.} by consideration of
things like mixing ratios in a pack of cards,
estimates of human propensity for various courses of behaviour, and so forth);
and then one multiplies each history's probability by the appropriate
influence factor ($C^{\sharp\mathrm{Higgses}}$, or a generalization thereof).
These {\em influenced but unnormalized} probabilities stand in the right
ratios to one another, but they do not in general sum to 1.
They must therefore be (globally) {\em normalized}
by dividing through by their sum.

This final global normalization is perhaps the least satisfying aspect
of the simplified approximate N\&N model.
However, it presents no technical difficulties,
other than the potential complexity of computing the sum.
We would be interested to know if a (presumably fully quantum)
version of their model can be found
which is free from the need for such normalization.

\section{The iterated card game: definition}
We described our iterated version of the N\&N card game
briefly in our introductory remarks.
Here we specify it precisely, define useful notation referring to it,
and clarify and motivate our assumptions
(which sometimes differ from N\&N's) about the various defined terms.

At the start of each time period, the intention of the human agents
enacting the protocol is to draw a card from a pack, and follow the
instruction printed on it, which will be either
``don't run LHC this period" or ``run LHC normally this period",
with mixing ratio $p : 1-p$.
We assume ``community self-discipline",
{\it i.e.} the instruction drawn is followed.
(Recall that the likely lack of such self-discipline in the one-shot version
was our motive for moving to an iterated version of the game.)

Neither the LHC as a piece of hardware, nor the society surrounding it,
offers a perfect guarantee that an intended run will actually happen.
In \cite{0707.1919}, N\&N use the blanket term ``accident" for
the thwarting of the successful running of LHC for whatever reason.
The most straightforward case would be failure of the hardware,
but they also give ``war between the member states of CERN" as an example
of the diversity of possible causes.

The potential causes of failure of LHC can broadly be divided into
what might be called ``endogenous" and ``exogenous" kinds.
An ``endogenous" failure is associated with the act of trying a run,
and basically refers to failure of the hardware.
For simplicity we take such a failure to be severe enough to
ruin LHC permanently,
although one could envisage a yet finer split into categories of damage
whose repair takes various lengths of time, and so forth.

An ``exogenous" failure is associated with the surrounding society,
and is independent of whether or not a run would be tried this time period.
Indeed, it is simplest to take exogenous failure as manifesting at the
very {\em beginning} of a period --
{\it i.e.} the agents do not even draw a card.
Analogously to our simplifying assumption about endogenous failure above,
we assume an exogenous failure brings the whole protocol to an end.
War between the member states of CERN would be the paradigm example.

N\&N use the term ``accident", assumed to have probability $a$,
to cover all failures of LHC, but for the iterated version we will
distinguish the endogenous and exogenous failures,
referring to them as ``accident" (to the hardware)
and ``breakdown" (of the whole protocol) respectively,
with respective probabilities (per opportunity) $a$ and $b$.
(These terms are chosen for their mnemonic alphabetical adjacency
and are not intended to be perfectly descriptive of the vast class of
conceiveable endogenous and exogenous failures.)

\subsection*{Event structure in one time period}
If the iterated game is still in progress as we enter a new time period
({\it i.e.} if no ``accident" or ``breakdown" has occurred earlier),
exactly one of the four available events $A$, $B$, $R$, $\bar{R}$
will occur in this period.
These stand for ``accident", ``breakdown",
``run of LHC", and ``no run of LHC" respectively.
The tree of alternatives is as follows.
Note that their listed probabilities are, for now, the straightforward
``uninfluenced" ones got by the usual sort of probabilistic reasoning.

\begin{itemize}
\item
First, the surrounding society impinges on the protocol.
With probability $b$ the protocol is exogenously brought to an end here
(``breakdown").
\\
{\bf Symbol $B$; (uninfluenced) probability $p_B \ideq b$.}

\item
With probability $1-b$ the protocol continues and a card is drawn
from the pack.
The card has instruction ``don't run LHC this period"
with probability $p$, in which case no run is attempted.
The protocol continues to the next time period.
\\
{\bf Symbol $\bar{R}$; (uninfluenced) probability $p_{\bar{R}} \ideq (1-b)p$.}

\item
Alternatively, with probability $1-p$, the card drawn has instruction
``run LHC normally this period", in which case a run is attempted.
At this point we have the opportunity for an ``accident" (endogenous failure)
to occur, with probability $a$. Like ``breakdown" this halts the protocol.
\\
{\bf Symbol $A$; (uninfluenced) probability $p_A \ideq (1-b)(1-p)a$.}

\item
If no accident occurs the run proceeds successfully.
(This is the event that will attract a multiplicative influence due to
the production of Higgs particles.)
The protocol continues to the next time period.
\\
{\bf Symbol $R$; (uninfluenced) probability $p_R \ideq (1-b)(1-p)(1-a)$.}
\end{itemize}

\subsection*{Classical histories of the iterated game}
The iterated card game with the above event structure and stopping rules
has classical histories of the form
$R\bar{R}\bar{R}R\bar{R}A$ or
$\bar{R}R\bar{R}\bar{R}RR\bar{R}RB$.
The pattern of $R$ and $\bar{R}$ in these examples is arbitrary;
a history can be any sequence of zero or more of $R$ or $\bar{R}$,
in any order and admixture,
followed by either $A$ or $B$ as terminating event.

It is not to be regarded as tragic that all histories end in
accident or breakdown -- the protocol is an expression of the sentiment
``if we {\em can} still run LHC, {\em do} keep running it"
(modulo the result of the card draw in each period of course).
We are interested in questions such as:
What happens {\em before} the iterated game inevitably ends?
How long does the game typically last, what is the pattern of events,
and which way does it end -- $A$ or $B$?

A history with $n_R$ $R$s, $n_{\bar{R}}$ $\bar{R}$s, $n_A$ $A$s, and $n_B$ $B$s
(these last two being either 1 and 0, or 0 and 1 of course)
has (uninfluenced) probability
\begin{eqnarray}\label{P_uninfl}
P_\mathrm{uninfl.}(\mathrm{history})
=
{p_R}^{n_R} {p_{\bar{R}}}^{n_{\bar{R}}} {p_A}^{n_A} {p_B}^{n_B}
\end{eqnarray}
It is automatic, and an easy exercise in algebra to verify,
that the sum of the value of this expression over all the available histories
is 1.

\subsection*{Multiplicative influence on a classical history}
Let us assume that a run of LHC which lasts one time period
produces some standard typical number of Higgs particles,
$\sharp\mathrm{Higgses\_in\_one\_run}$,
and let us write $\sigma$ for the expression
$C^{\sharp\mathrm{Higgses\_in\_one\_run}}$.
(Note that we have $\sigma < 1$ since $C < 1$.)
Then a history with $n_R$ $R$s will be subject to a multiplicative influence
of $\sigma^{n_R}$.

For conceptual clarity we may as well ``distribute" this influence down to
the probabilities of the events which contribute to it.
If for each event type $E \in \{A, B, R, \bar{R}\}$
we write $\phi_E$ for the {\em influenced}
(but still unnormalized) probability of that event type
($p_E$ being its uninfluenced probability, as listed earlier),
then we have $\phi_A = p_A$, $\phi_B = p_B$, $\phi_{\bar{R}} = p_{\bar{R}}$,
but $\phi_R = \sigma p_R = (1-b)(1-p)(1-a)\sigma$.

A history with $n_R$ $R$s, $n_{\bar{R}}$ $\bar{R}$s, $n_A$ $A$s, and $n_B$ $B$s
can thus have its influenced but unnormalized probability written as
\begin{eqnarray}\label{P_infl_unnorm}
P_\mathrm{infl., unnorm.}(\mathrm{history})
=
{\phi_R}^{n_R} {\phi_{\bar{R}}}^{n_{\bar{R}}} {\phi_A}^{n_A} {\phi_B}^{n_B}
\end{eqnarray}
Of course the sum-over-histories of {\em this} expression
will be less than 1.

\subsection*{Some brief notes on where our assumptions differ from N\&N's}
We already noted above our splitting of N\&N's single failure category
into separate categories of endogenous ``accident" and exogenous ``breakdown".
For the benefit of readers steeped in N\&N's notation and assumptions,
we briefly note other differences between our setup and theirs.

\begin{itemize}
\item
In their quantitative analysis N\&N assume for simplicity
that any branch of their tree of alternatives with Higgs particle production
can be dropped altogether in their model
({\it i.e.} can be taken as having influenced probability zero, or negligible).
In our notation this is like taking $\sigma << 1$.
This would not be appropriate for the iterated protocol,
since after all the time periods could be very short
(minutes, seconds...)\footnote{
	At least for conceivable accelerator hardware.
	We leave aside whether ramping up and down the {\em actual} hardware
	of LHC on short timescales would be at all feasible!
}
and not much multiplicative influence will creep into any one period.
Histories with copious Higgs production are of course still strongly
suppressed, but the mathematical expression of this fact becomes,
not that $\sigma$ be small, but that it be raised to a high power ($n_R$).

\item
N\&N take the pack mixing ratio $p$, and their accident probability $a$
(which we disaggregate into $a$ and $b$), to be small ($<< 1$).
Their motive for choosing $p$ small is the unacceptability of the card game
to the community if $p$ is large and the associated instruction is
``shut down LHC forever".
The iterated protocol does not have this problem
(at least not to that stark extent!) and we can choose $p$ more freely.
As for $a$ and $b$ (or for N\&N just $a$),
they assume smallness to keep rough comparability with $p$
(or just to simplify the mathematics).
We prefer to drop that assumption too, since after all
complex machinery {\em does}
have a non-negligible accident rate, and societies {\em do} often undergo
convulsions leading to the interruption of big projects
(war being only one of many possible scenarios here).
Thus in our analysis we do not assume $p$, $a$ or $b$ small.
We are of course free to plug in small values to any expressions derived,
but we derive the expressions without approximation.

\item
N\&N perform a kind of ``meta-analysis" by introducing a parameter
$r$ to represent an externally agreed probability (like a Bayesian prior)
for their model being correct at all.
(They also very modestly set $r<<1$, in effect taking the meta-analyst to be
a natural skeptic of new theories.)

In this paper we are always asking:
what would the world be like if the N\&N model were true?
Thus in effect we are always setting $r=1$, at least for the purposes
of answering that overarching question.
Note, though, that with our notation one can ``switch off" the model
by setting $\sigma=1$. Therefore, one {\em could} perform a meta-analysis by
giving a prior distribution for $\sigma$
({\it i.e.} $1-r$ for $\sigma=1$,
and $r$ shared out among values of $\sigma<1$).
But this is not something that we do in this paper.

\item
Finally, N\&N perform a cost-benefit analysis by assigning equivalent
money or utility values to the outcomes of various scenarios.
(Indeed, this is their main motive for having a meta-analysis variable $r$.)
We will not do this explicitly, although from time to time we will
refer to the general intuitive ``goodness" or ``badness"
of various histories, from the viewpoint both of the scientific community
(awaiting run data from LHC) and of the broader society
(presumably hoping to avoid at least the more convulsive forms of the
category ``breakdown").
\end{itemize}

\section{The iterated card game: analysis}
To analyse the iterated card game we must compute the {\em normalized}
influenced probabilities of the available histories.
We first compute the normalization factor $N.F.$ by
summing the influenced but unnormalized probabilities of all histories:

\begin{eqnarray}\label{sum_P_infl_unnorm}
N.F.
\ideq
\sum_{\mathrm{histories}} P_\mathrm{infl., unnorm.}(\mathrm{history})
&=&
\sum_{0 \leq n_{R|{\bar{R}}} < \infty}
	{\phi_{R|{\bar{R}}}}^{n_{R|{\bar{R}}}} \phi_{A|B}
\cr \cr &=&
\frac {\phi_{A|B}} {1 - \phi_{R|{\bar{R}}}}
\end{eqnarray}
\\
(In (\ref{sum_P_infl_unnorm}) and elsewhere
we abbreviate sums of probabilities or occurrence counts over alternatives
by listing the alternatives in bar-separated fashion -- that is,
$n_{R|{\bar{R}}}$ means $n_R + n_{\bar{R}}$, and so forth.)

We can then divide the earlier expression (\ref{P_infl_unnorm})
by $N.F.$ to obtain the normalized influenced probability for a history with
$n_R$ $R$s, $n_{\bar{R}}$ $\bar{R}$s, $n_A$ $A$s, and $n_B$ $B$s:

\begin{eqnarray}\label{P_infl_norm}
P_\mathrm{infl., norm.}(\mathrm{history})
\ideq
\frac
  {P_\mathrm{infl., unnorm.}(\mathrm{history})}
  {N.F.}
=
\frac
  {{\phi_R}^{n_R} {\phi_{\bar{R}}}^{n_{\bar{R}}} {\phi_A}^{n_A} {\phi_B}^{n_B}}
  {\frac {\phi_{A|B}} {1 - \phi_{R|{\bar{R}}}}}
\end{eqnarray}
\\
Let us specialize this rather forbidding expression to a history ending
in $A$ or $B$ respectively -- {\it i.e.} with $n_A = 1$, $n_B = 0$ or
$n_A = 0$, $n_B = 1$ respectively.

\begin{eqnarray}
P_\mathrm{infl., norm.}(\mathrm{history~ending~in~A})
&=&
{\phi_R}^{n_R} {\phi_{\bar{R}}}^{n_{\bar{R}}}
(\frac{\phi_A}{\phi_{A|B}} [1 - \phi_{R|{\bar{R}}}])
\end{eqnarray}
\begin{eqnarray}
P_\mathrm{infl., norm.}(\mathrm{history~ending~in~B})
&=&
{\phi_R}^{n_R} {\phi_{\bar{R}}}^{n_{\bar{R}}}
(\frac{\phi_B}{\phi_{A|B}} [1 - \phi_{R|{\bar{R}}}])
\end{eqnarray}
\\
This is the {\em same} probability distribution
(assignment of probabilities to histories) as the
{\em uninfluenced} distribution we would have obtained if,
instead of $p_R$, $p_{\bar{R}}$, $p_A$, $p_B$ in (\ref{P_uninfl}) above,
we had used $\rho_R$, $\rho_{\bar{R}}$, $\rho_A$, $\rho_B$,
defined as follows:

\begin{eqnarray}
\rho_R \ideq \phi_R = \sigma p_R = (1-b)(1-p)(1-a)\sigma
\end{eqnarray}
\begin{eqnarray}
\rho_{\bar{R}} \ideq \phi_{\bar{R}} = p_{\bar{R}} = (1-b)p
\end{eqnarray}
\begin{eqnarray}
\rho_A &\ideq& \frac{\phi_A}{\phi_{A|B}} [1 - \rho_{R|{\bar{R}}}]
       = \frac{\phi_A}{\phi_{A|B}} [1 - \phi_{R|{\bar{R}}}]
=
\frac{p_A}{p_{A|B}} [1 - \sigma p_R - p_{\bar{R}}]
\cr \cr &=&
\frac{(1-b)(1-p)a}{(1-b)(1-p)a + b} [1 - (1-b)(1-p)(1-a)\sigma - (1-b)p]
\end{eqnarray}
\begin{eqnarray}
\rho_B &\ideq& \frac{\phi_B}{\phi_{A|B}} [1 - \rho_{R|{\bar{R}}}]
       = \frac{\phi_B}{\phi_{A|B}} [1 - \phi_{R|{\bar{R}}}]
=
\frac{p_B}{p_{A|B}} [1 - \sigma p_R - p_{\bar{R}}]
\cr \cr &=&
\frac{b}{(1-b)(1-p)a + b} [1 - (1-b)(1-p)(1-a)\sigma - (1-b)p]
\end{eqnarray}
\\
That is, with $\{\rho_E\}$ ($E \in \{A, B, R, \bar{R}\}$) defined as above,
the normalized influenced probability of every history is given by
\begin{eqnarray}
P_\mathrm{infl., norm.}(\mathrm{history})
=
{\rho_R}^{n_R} {\rho_{\bar{R}}}^{n_{\bar{R}}} {\rho_A}^{n_A} {\rho_B}^{n_B}
\end{eqnarray}

Thus the $\{\rho_E\}$ can be thought of as
{\em de facto} normalized influenced probabilities for the event types $E$.
Note, however, that for more complicated protocols
 -- for example involving agents with memory, who do things like
adjust the mixing ratio $p$ depending on the outcomes of previous draws --
the normalized influenced probability distribution is unlikely to be
a mere ``re-parametrization" of the uninfluenced one.
That is, there will {\bf not}, in general, exist a choice of
$\{\rho_E\}$ or analogues thereof
such that substituting them for the original $\{p_E\}$ or analogues thereof
mimics the effects of N\&N-style backward causation.
We just got lucky with the simple memoryless protocol explored here!

A helpful intuitive story for obtaining
$\rho_R$, $\rho_{\bar{R}}$, $\rho_A$, $\rho_B$
from $p_R$, $p_{\bar{R}}$, $p_A$, $p_B$
goes like this:
First, the probabilities for the non-terminating events ($R$, $\bar{R}$)
shrink under whatever multiplicative influence (if any)
they are individually subject to, which happens to be an
{\em unbalanced} (non-ratio-preserving) shrinkage,
since $\rho_R < p_R$ but $\rho_{\bar{R}} = p_{\bar{R}}$.
Then, the probabilities for the terminating events ($A$, $B$)
grow in {\em balanced} (ratio-preserving) style
 -- {\it i.e.} the ratio $\rho_A : \rho_B$ is the same as $p_A : p_B$ --
to the extent necessary to ``fill the gap" and ensure the four
event probabilities sum to 1.

With the normalized influenced probability distribution established,
we can proceed to study
the effects of N\&N-style backward causation on
the statistics of the iterated card game.

\subsection*{The physics community perspective: how much run data can we
		squeeze out of LHC?}
Let us first look at things from the perspective of a caricatured
LHC physics community, which we define as {\em not} caring
about goings-on in the broader society, but instead concerning itself
only with how much run data can be squeezed out of LHC
before the game ends (in ``accident" or ``breakdown").

The goal from this perspective is to maximize $\expectation{n_R}$,
the expectation value of the number of periods a successful run of LHC
takes place.
We have $a$, $b$, $\sigma$ fixed by the properties of
the LHC hardware, the surrounding society, and the N\&N Lagrangian
respectively, but $p$ choosable. We compute $\expectation{n_R}$:

\begin{eqnarray}\label{expectation_n_R}
\expectation{n_R}
&\ideq&
\sum_{\mathrm{histories}}
	P_\mathrm{infl., norm.}(\mathrm{history}) n_R(\mathrm{history})
\cr \cr &=&
\frac{\rho_R}{\rho_{R|{\bar{R}}}}
\sum_{\mathrm{histories}}
	P_\mathrm{infl., norm.}(\mathrm{history})
	n_{R|{\bar{R}}}(\mathrm{history})
\cr \cr &=&
\frac{\rho_R}{\rho_{R|{\bar{R}}}}
\sum_{0 \leq n_{R|{\bar{R}}} < \infty}
	[{\rho_{R|{\bar{R}}}}^{n_{R|{\bar{R}}}} \rho_{A|B}]
	[n_{R|{\bar{R}}}]
\cr \cr &=&
\frac {\rho_R} {\rho_{A|B}}
=
\frac {(1-b)(1-p)(1-a)\sigma} {1 - (1-b)(1-p)(1-a)\sigma - (1-b)p}
\end{eqnarray}
\\
By differentiating w.r.t. $p$ we find that this expression is
monotonically decreasing in $p$ in the range $0 \leq p \leq 1$.
Thus the physics community would want to set $p=0$,
that is to say, not play the card game at all
but just run LHC in every period.
With this choice the expression for $\expectation{n_R}$ becomes:
\begin{eqnarray}\label{expectation_n_R_p=0}
\expectation{n_R}(p=0)
=
\frac {(1-b)(1-a)\sigma} {1 - (1-b)(1-a)\sigma}
\end{eqnarray}

As expected this goes to zero as $\sigma \to 0$.
However, it goes to zero {\em more slowly}
than would any $\expectation{n_R}(p>0)$.

\subsection*{The broader society perspective: how long till breakdown?}
We now turn to the perspective of the ``broader society".
Our caricature here is the opposite of that for the physics community.
The broader society is defined as having no interest in the
quantity of run data from LHC, nor in whether it suffers hardware failure
(our category ``accident").
Rather, its concern is with our category ``breakdown"
 -- this can be presumed, at least in its more convulsive forms,
to be an unpleasant experience to live through.

Recall that $B$ has an exogenously given risk rate $b$ in our simple setup.
That is, even in the absence of LHC, or more pertinently,
even after an accident ($A$) has halted the card game,
$B$ events continue to occur with probability $b$ per time period.
(Of course, by ``$B$" in such non-game or post-game circumstances
we mean whatever sort of event in society {\em would} halt LHC runs
had there been any. The broader society dislikes such events
not for their actual or counterfactual halting of LHC,
but for their potentially convulsive nature generally.)
Thus we should not ask: Can we avoid $B$ altogether?
but rather: What is the expected time to the next occurrence of $B$?
In the absence of influence from the future this is $1/b$,
by elementary properties of the exponential distribution.
We now compute it for the influenced probability distribution
given by the $\{\rho_E\}$.

For a history ending in $B$, the expected time of occurrence of $B$
(numbering the periods $1,2,3...$)
is simply the {\em actual} time $B$ occurs,
{\it i.e.} the length of that history:

\begin{eqnarray}
\expectation{t_B}(\mathrm{history~ending~in~B})
&=&
t_B(\mathrm{history})
=
\mathrm{length}(\mathrm{history})
\end{eqnarray}
\\
For a history ending in $A$, we have no $B$ within the history,
but as discussed above we expect a post-game $B$ to occur eventually.
Once the accident has occurred
there is no influence from the future to contend with,
{\it i.e.} the straightforward uninfluenced risk rate $b$ applies.
Thus an occurrence of $A$ at time $t_A$
can be taken to herald an occurrence of $B$
at (expected) later time $t_A + 1/b$:

\begin{eqnarray}
\expectation{t_B}(\mathrm{history~ending~in~A})
&=&
t_A(\mathrm{history}) + 1/b
\cr &=&
\mathrm{length}(\mathrm{history}) + 1/b
\end{eqnarray}
\\
We can now compute $\expectation{t_B}$:

\begin{eqnarray}\label{expectation_t_B}
\expectation{t_B}
&\ideq&
\sum_{\mathrm{histories}}
	P_\mathrm{infl., norm.}(\mathrm{history})
	\expectation{t_B}(\mathrm{history})
\cr \cr &=&
\sum_{\mathrm{histories~ending~in~B}}
	[P_\mathrm{infl., norm.}(\mathrm{history})]
	[\mathrm{length}(\mathrm{history})]
\cr \cr &+&
\sum_{\mathrm{histories~ending~in~A}}
	[P_\mathrm{infl., norm.}(\mathrm{history})]
	[\mathrm{length}(\mathrm{history}) + 1/b]
\cr \cr &=&
\sum_{0 \leq n_{R|{\bar{R}}} < \infty}
	[{\rho_{R|{\bar{R}}}}^{n_{R|{\bar{R}}}} \rho_B]
	[n_{R|{\bar{R}}} + 1]
+
\sum_{0 \leq n_{R|{\bar{R}}} < \infty}
	[{\rho_{R|{\bar{R}}}}^{n_{R|{\bar{R}}}} \rho_A]
	[n_{R|{\bar{R}}} + 1 + 1/b]
\cr \cr &=&
\frac{1}{p_B} - \frac{1}{p_{A|B}} + \frac{1}{\rho_{A|B}}
\cr \cr &=&
\frac{1}{b}
-
\frac{1}{(1-b)(1-p)a + b}
+
\frac{1}{1 - (1-b)(1-p)(1-a)\sigma - (1-b)p}
\end{eqnarray}
\\
(We omit the tedious algebra leading to the final
``sum or difference of reciprocals" form,
which seems to be the simplest possible.)

One can get an intuitive grip on (\ref{expectation_t_B}) by
setting $q \ideq 1-p$ and defining ``influence constants" $i_1, i_2$
as follows:
\begin{eqnarray}
i_1 &\ideq& (1-b)a
\cr
i_2 &\ideq& (1-b)(1-a)(1-\sigma)
\end{eqnarray}

Then (\ref{expectation_t_B}) becomes:
\begin{eqnarray}\label{expectation_t_B_intuitive}
\expectation{t_B} = \frac{1}{b} - \frac{1}{b + i_1q} + \frac{1}{b + (i_1+i_2)q}
\end{eqnarray}

The variation of $\expectation{t_B}$ with $q$ now becomes clear:
the three reciprocals are equal when $q=0$,
but as we increase $q$ they separate -- the first staying fixed,
the second and third falling, with the third (being added)
always smaller than the second (being subtracted).
An immediate consequence is that
$\expectation{t_B}$ is at its maximum when $q=0$ ({\it i.e.} when $p=1$).
However it is not in general
monotonically decreasing in $q$ in the range $0 \leq q \leq 1$:
differentiating w.r.t. $q$ shows it to reach a minimum at
$q = b/\sqrt{i_1(i_1+i_2)}$
(if that value is in the range $0..1$) and start increasing again,
though it never again attains its $q=0$ value.

We see then that the broader society would prefer to set $q=0$ ($p=1$),
that is to say, not run LHC at all.
Of course $\expectation{t_B}$ is then just $1/b$.
The nonmonotonicity of $\expectation{t_B}$ with $q$
deserves further comment, however.
It is best thought of as arising from a kind of ``cross-talk"
between, on the one hand, the effect of the influence from the future on
the timescale for halting the protocol
by {\em either} terminating event ($A$ or $B$),
and on the other,
the impact of $q$ on {\em which} terminating event it shall be.
The expected time for LHC to be halted for whatever reason is

\begin{eqnarray}\label{expectation_t_A|B}
\expectation{t_{A|B}}
&=&
\sum_{\mathrm{histories}}
	[P_\mathrm{infl., norm.}(\mathrm{history})]
	[\mathrm{length}(\mathrm{history})]
\cr \cr &=&
\sum_{0 \leq n_{R|{\bar{R}}} < \infty}
	[{\rho_{R|{\bar{R}}}}^{n_{R|{\bar{R}}}} \rho_{A|B}]
	[n_{R|{\bar{R}}} + 1]
\cr \cr &=&
\frac{1}{\rho_{A|B}}
=
\frac{1}{b + (i_1+i_2)q}
\end{eqnarray}
\\
-- and {\em this} is clearly monotonically decreasing in $q$.
On the other hand, the odds ratio ``ending in $A$" : ``ending in $B$"
is just $\rho_A : \rho_B$,
which we earlier observed is the same as $p_A : p_B$,
namely $(1-b)qa : b$,
which becomes more favourable to $A$ at the expense of $B$ as $q$ increases.
This provides a ``safety valve" effect: at high enough $q$
the protocol is indeed brought to a halt very quickly,
but with odds shifted in the direction of $A$,
which (after it happens) relaxes the risk of $B$ to its
uninfluenced level (timescale $1/b$).

\subsection*{A numerical example}
Let us put some numerical flesh on these algebraic bones.
Take the time period to be a week, and set
$a=b=0.0002$ (once-per-century uninfluenced accident or breakdown timescale),
$\sigma=0.9$.
If we choose $q=0$ we have
$\expectation{t_B} = \expectation{t_{A|B}} = 1/b = 5000$,
the uninfluenced timescale of a century or so.

If we increase $q$ to 0.04
(near the minimum-$\expectation{t_B}$ value, which in this example
is $q \approx 0.0447$),
we obtain
$\expectation{t_{A|B}} \approx 238$.
Thus already the likely timescale for {\em something} to halt the protocol
has shrunk from a century to a few years.
Furthermore, the odds $\rho_A : \rho_B$ are about $1:25$,
{\it i.e.} the reason will almost certainly turn out to be ``breakdown".
These effects combine to yield $\expectation{t_B} \approx 430$.

If we make the choice $q=1$, we get $\expectation{t_{A|B}} \approx 10$:
something will halt the protocol in a matter of weeks.
However, the odds $\rho_A : \rho_B$ are now $\approx 1:1$,
so we have the ``safety valve" of a 0.5 probability that
the reason will turn out to be not ``breakdown" but ``accident".
This is reflected in the value $\expectation{t_B} \approx 2510$.
In effect we are tossing a fair coin and gambling on breakdown in weeks
versus a century.

\section{Statistical cosmic censorship}
The above analysis and numerical example
shows that N\&N-style backward causation can be pretty powerful stuff.
It can greatly enhance the likelihood of a quick occurrence of
what we would normally regard as remote contingencies.
It is natural to ask: what are the prospects for {\em empirical} study
of the influence from the future? Would we know it when we see it?

If we enact the iterated card game protocol and quickly experience
an event of the sort we have called ``accident" or ``breakdown",
we are left in the awkward situation of not really knowing
(as opposed to estimating) the {\em uninfluenced} probabilities of
these contingencies, which depend on the nature of complex hardware
and a complex society.
People of goodwill will disagree in their estimates.
Those whose estimates are on the high side
will simply give a resigned shrug.
``That's life", they will say.

There {\em is}, however, one feature of the protocol which has a clear-cut
uninfluenced probabilistic structure:
the drawing of a card in each time period.
Everyone can agree that the uninfluenced probability of drawing
the instruction ``don't run LHC this period" is $p$.
And as we mentioned in our introductory remarks,
the pattern of runs and idle periods is the protocol's ``deliverable",
available for statistical scrutiny by interested parties.
What can they conclude if they perform a statistical analysis?
Can they detect the hand of N\&N-style backward causation?

A card draw occurs part-way through its time period and there is not a
one-one mapping between results of draws and event labels.
A drawing of ``run LHC normally this period"
(which we will abbreviate ``yes", $Y$)
happens in events $A$ and $R$.
A drawing of ``don't run LHC this period"
(which we will abbreviate ``no", $N$)
happens only in event $\bar{R}$.
No draw at all occurs in event $B$ since exogenous halting
is assumed to occur at the beginning of a time period.
Hence a draw pattern (of $Y$ and $N$) is got from a regular event history
by dropping $B$, changing $A$ and $R$ into $Y$,
and changing $\bar{R}$ into $N$.

We can conveniently quantify the strength of the influence from the future
on the pattern of card draws
by defining the {\em discrepancy} $D$ between the actual number of
drawings of ``no" and the number expected purely on the basis of
multiplying the total number of draws by the ``no" mixing ratio $p$.
That is:
\begin{eqnarray}\label{discrepancy}
D \ideq n_N - (n_{Y|N})p
\end{eqnarray}

Those using uninfluenced probabilistic reasoning will assign this an
expectation value $\expectation{D}=0$ --
although of course they anticipate that the actual (outcome) discrepancy
will suffer binomial-style fluctuation of order $\sqrt{(n_{Y|N})pq}$
about its zero expected value.

We compute the expected discrepancy under the {\em influenced}
probability distribution given by the $\{\rho_E\}$:

\begin{eqnarray}\label{expectation_discrepancy}
\expectation{D}
&\ideq&
\sum_{\mathrm{histories}}
	P_\mathrm{infl., norm.}(\mathrm{history})
	D(\mathrm{history})
\cr \cr &=&
\sum_{\mathrm{histories}}
	P_\mathrm{infl., norm.}(\mathrm{history})
	[n_N - (n_{Y|N})p](\mathrm{history})
\cr \cr &=&
\sum_{\mathrm{histories}}
	P_\mathrm{infl., norm.}(\mathrm{history})
	[n_{\bar{R}} - (n_{A|R|\bar{R}})p](\mathrm{history})
\cr \cr &=&
\sum_{0 \leq n_{R|{\bar{R}}} < \infty}
	[{\rho_{R|{\bar{R}}}}^{n_{R|{\bar{R}}}} \rho_A]
	[\frac{\rho_{\bar{R}}}{\rho_{R|{\bar{R}}}} n_{R|{\bar{R}}}
		-
	 (n_{R|{\bar{R}}} + 1)p]
\cr \cr &+&
\sum_{0 \leq n_{R|{\bar{R}}} < \infty}
	[{\rho_{R|{\bar{R}}}}^{n_{R|{\bar{R}}}} \rho_B]
	[\frac{\rho_{\bar{R}}}{\rho_{R|{\bar{R}}}} n_{R|{\bar{R}}}
		-
	 (n_{R|{\bar{R}}})p]
\cr \cr &=&
[~p~]	~
[~\frac{p_B}{p_{A|B}}~]	~
[~1 - \frac{p_{A|B}}{\rho_{A|B}}~]
\end{eqnarray}
\\
(We again omit the tedious algebra leading to the final form.)
$\expectation{D}$ is thus the product of three factors
each manifestly in the range $0..1$,
and is therefore in the range $0..1$ also.\footnote{
	It may be helpful to make judicious use of $q \ideq 1-p$
	and the ``influence constants" $i_1, i_2$ defined earlier:
	\begin{eqnarray}
	\expectation{D}
	=
	[~p~]	~
	[~\frac{b}{b + i_1q}~]	~
	[~\frac{i_2q}{b + (i_1+i_2)q}~]
	\end{eqnarray}
	It is then immediate that each factor is in the range $0..1$.
}
In other words, the influence from the future
shifts $\expectation{D}$ by
{\bf less than a single card}
from its uninfluenced value --
a shift lost in the order-$\sqrt{(n_{Y|N})pq}$ fluctuation in
the actual (outcome) discrepancy.
Statistical analysis of the card draws will {\em not} reveal
the hand of N\&N-style backward causation.

Hence we find ourselves in a startling epistemic situation
if we try to do statistical analysis in an N\&N world.
The ``epistemically opaque" aspects of the enactment of the protocol,
such as the timescale for the one-off event of being forcibly halted,
are potentially greatly influenced by N\&N-style backward causation;
while the one clearly epistemically accessible aspect
 -- the drawing of a card, which has a straightforward
probability structure controlled by the mix of cards in the pack,
and which is a repeated rather than one-off event,
allowing the accumulation of statistics --
is not {\em detectably} influenced at all!
It is thus peculiarly difficult (at least with this protocol)
for the inhabitants of a world subject to N\&N-style backward causation
to gather evidence revealing their predicament.
This situation might be called ``statistical cosmic censorship".

We close by conjecturing that statistical cosmic censorship
may be a generic, or at least common, feature of models with
N\&N-style backward causation -- and, perhaps,
of theories with other types of non-standard casual structure too.
We would welcome efforts to define the syndrome more precisely,
and to explore what reliable knowledge
can and cannot be acquired by agents
enacting experimental protocols of their choice
in a world where the future, as well as the past, informs their actions.

\section*{Acknowledgements}
%%% I.S. would like to thank
I would like to thank
Paulo Pires-Pacheco for drawing my attention to \cite{0707.1919},
and for robust discussion of it and other things;
%%% and Shashank Virmani,
and Adilah Hussein and Shashank Virmani,
for discussions on causality in physics and much else besides.

\end{document}